\documentclass[12pt,a4paper]{article}

\usepackage[top=1in, bottom=1.25in, left=1in, right=1in]{geometry}
\usepackage{lineno}  
\usepackage{xspace} 
\usepackage{caption}

\usepackage{graphicx}  
\usepackage{color}
\usepackage{colortbl}
\graphicspath{{./figs/}} 

\usepackage{ifthen} 
\newboolean{articletitles}
\setboolean{articletitles}{true} 

\usepackage{amsmath} 
\usepackage{amssymb}
\usepackage{amsfonts}
\usepackage{multirow}

\def\papertitle{Projections for Key Measurements in Heavy Flavour Physics} 
\def\paperasciititle{Projections for Key Measurements in Heavy Flavour Physics} 
\def\paperauthors{%
F.~Blanc, A.~Gaz, T.~Gershon, J.~Libby, R.~Novotny, M.~Pierini, C.~Rovelli, S.~Turchikhin%
}

\usepackage[pdftex,
            pdfauthor={\paperauthors},
            pdftitle={\paperasciititle}]{hyperref}
\usepackage{hyperxmp}

\usepackage[bottom,flushmargin,hang,multiple]{footmisc}
\usepackage[all]{hypcap} 
\usepackage{comment}

\def\CP                {{\ensuremath{C\!P}}\xspace}
\def\jpsi     {{\ensuremath{{J\mskip -3mu/\mskip -2mu\psi}}}\xspace}
\newcommand{\eg}{\mbox{\itshape e.g.}\xspace}

\newcommand{\tev}{\text{Te\kern -0.1em V}\xspace}
\newcommand{\gev}{\text{Ge\kern -0.1em V}\xspace}
\newcommand{\mev}{\text{Me\kern -0.1em V}\xspace}
\newcommand{\kev}{\text{ke\kern -0.1em V}\xspace}
\newcommand{\ev}{\text{e\kern -0.1em V}\xspace}
 
\newcommand{\mevc}{\ensuremath{\text{Me\kern -0.1em V\!/}c}\xspace}
\newcommand{\gevc}{\ensuremath{\text{Ge\kern -0.1em V\!/}c}\xspace}
\newcommand{\mevcc}{\ensuremath{\text{Me\kern -0.1em V\!/}c^2}\xspace}
\newcommand{\gevcc}{\ensuremath{\text{Ge\kern -0.1em V\!/}c^2}\xspace}

\def\cm   {\text{cm}\xspace}

\def\fb   {\ensuremath{\text{fb}}\xspace}
\def\invfb   {\ensuremath{\fb^{-1}}\xspace}

\def\sec  {\ensuremath{\text{s}}\xspace}

\def\ps   {\ensuremath{\text{ps}}\xspace}

\def\mhz  {\ensuremath{\text{MHz}}\xspace}

\def\degrees{\ensuremath{^\circ}\xspace}

\usepackage{cite} 
\usepackage{mciteplus}

\begin{document}

\renewcommand{\thefootnote}{\fnsymbol{footnote}}
\setcounter{footnote}{1}

\begin{titlepage}
\pagenumbering{roman}

\begin{minipage}{0.22\textwidth}
    \includegraphics[width=\textwidth]{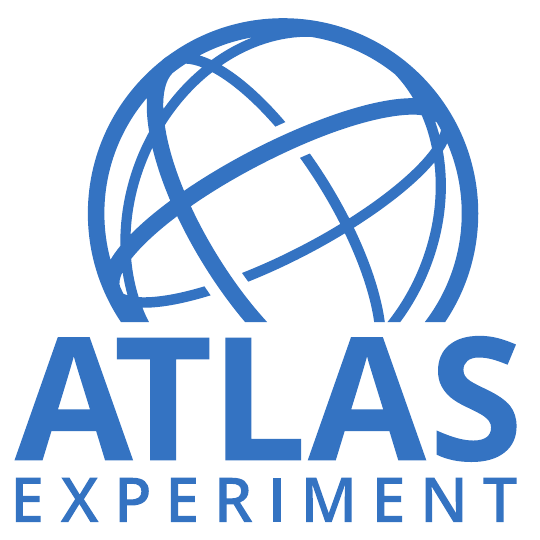}
\end{minipage}
\hfill
\begin{minipage}{0.22\textwidth}
    \includegraphics[width=\textwidth]{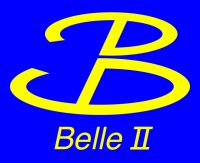}
\end{minipage}
\hfill
\begin{minipage}{0.20\textwidth}
    \includegraphics[width=\textwidth]{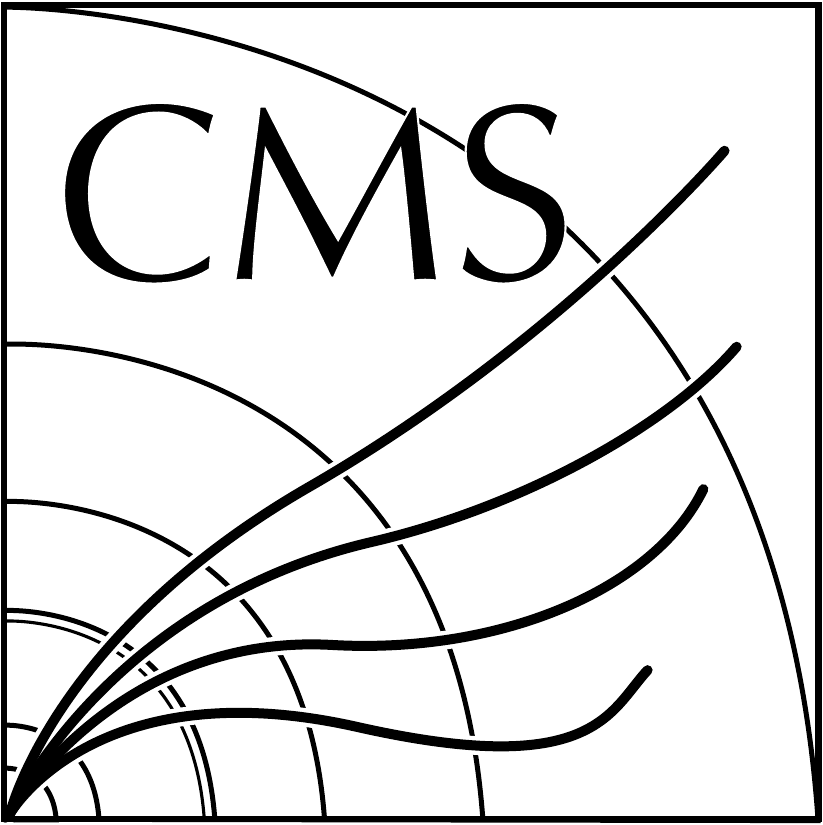}
\end{minipage}
\hfill
\begin{minipage}{0.24\textwidth}
    \includegraphics[width=\textwidth]{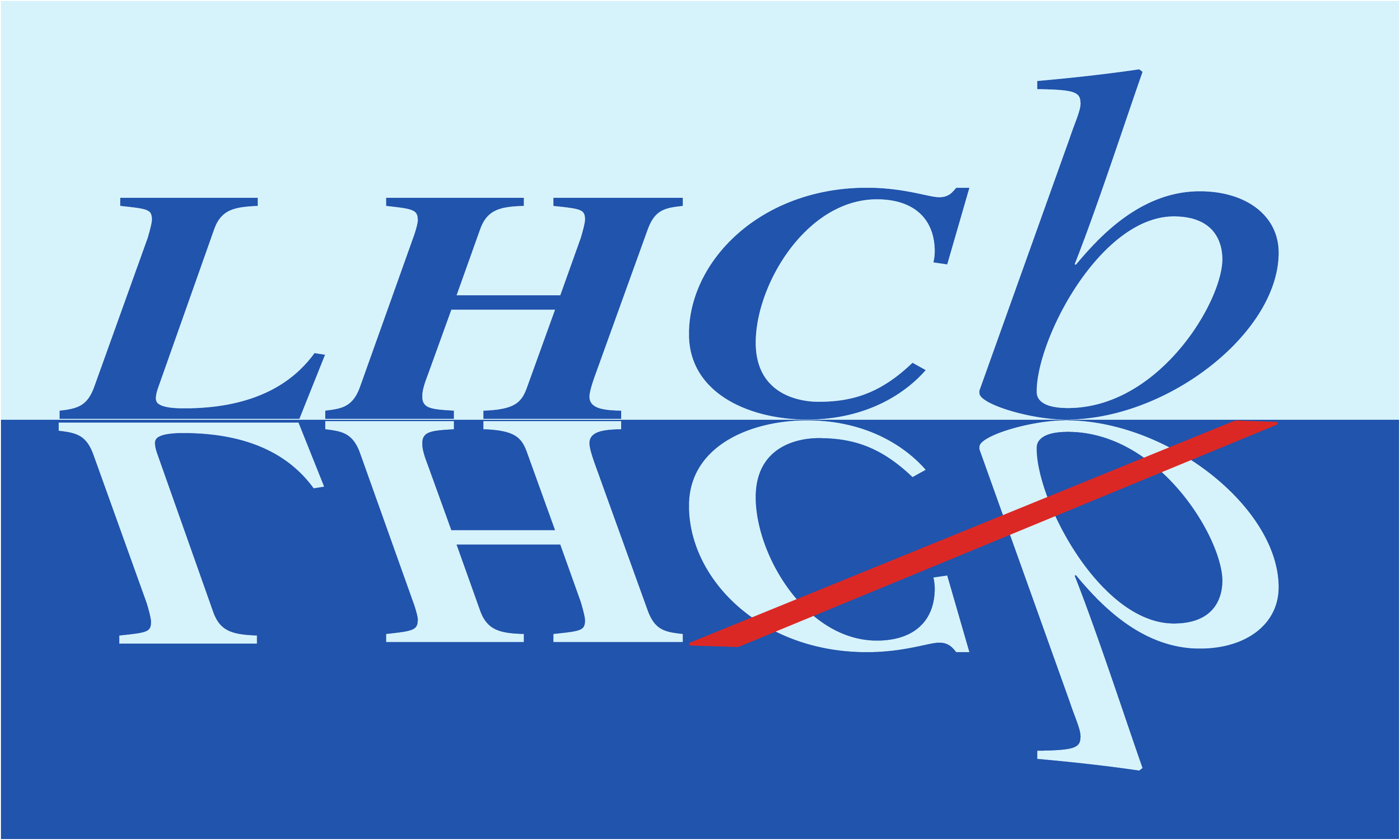}
\end{minipage}

\begin{flushright}
    ATL-PHYS-PUB-2025-020, CMS-BPH-25-001, LHCb-PUB-2025-008, \\
    Belle II Preprint 2025-007, KEK Preprint 2025-5\\[2ex]
    {\today}
\end{flushright}

\vspace*{0.7cm}

{\normalfont\bfseries\boldmath\huge 
\begin{center}
    \papertitle \\
    \vspace*{0.5cm}
  {\normalsize Input to the European Particle Physics Strategy Update 2024--26}
\end{center}
}

\vspace*{0.7cm}

\begin{center}
    The ATLAS, Belle~II, CMS and LHCb collaborations%
\end{center}

\vspace{\fill}

\begin{abstract}
    \noindent
    Precision studies of flavour-changing processes involving quarks and leptons provide a number of ways to improve knowledge of the Standard Model and search for physics beyond it.
    There are excellent short- and mid-term prospects for significantly improved measurements in heavy flavour physics (involving $b$ and $c$ hadrons and $\tau$ leptons), with upgrades in progress or planned for the ATLAS, CMS and LHCb experiments exploiting proton-proton collisions at CERN's Large Hadron Collider, and for the Belle~II experiment operating with electron-positron collisions from the SuperKEKB accelerator in KEK. 
    The expected sensitivities that can be achieved from these experiments for a number of key observables are presented, highlighting the complementarity of the different experiments and showing how the precision will improve with time.
    This international programme in heavy flavour physics will result in unprecedented capability to probe this sector of the Standard Model and, potentially, observe imprints of physics at higher energy scales than can be accessed directly. 
\end{abstract}

\vspace{\fill}

\end{titlepage}

\newpage
\setcounter{page}{2}
\mbox{~}

\cleardoublepage
\pagestyle{plain} 
\setcounter{page}{1}
\pagenumbering{arabic}
\renewcommand{\thefootnote}{\arabic{footnote}}
\setcounter{footnote}{0}


\section{Introduction}
\label{sec:introduction}

The flavour sector of the Standard Model (SM) remains enigmatic.  
There is no fundamental understanding of why so many different types of quarks and leptons exist, nor why their masses and mixings have such striking patterns.  
Moreover, while the SM provides a source of \CP\ violation through the Cabibbo--Kobayashi--Maskawa (CKM) quark mixing matrix~\cite{Cabibbo:1963yz,Kobayashi:1973fv}, and fulfils the Sakharov conditions necessary to generate a matter-dominated Universe~\cite{Sakharov:1967dj}, the SM is insufficient to produce the matter-antimatter asymmetry that is observed around us.
Resolving this problem is among the top priorities of contemporary physics.

In addition, a number of features of SM electroweak interactions result in the flavour sector being highly sensitive to potential contributions from physics beyond the SM.  
These include the fact that the CKM matrix has only four free parameters, which describe phenomena related to quark-flavour changes over a broad range of energy scales from nuclear beta decay to single top quark production.  
One of these parameters is the sole source of \CP violation in the SM~\cite{Kobayashi:1973fv}, from which all \CP-violating observables can in principle be predicted, allowing many different searches for effects that are inconsistent with these predictions.
Similarly, the suppression of flavour-changing neutral currents in the SM~\cite{Glashow:1970gm} may not be present in extended models.
Since the SM contributions to loop processes are small in the quark sector (\eg\ $b \to s\gamma$) and practically absent in the lepton sector (\eg\ $\tau \to \mu\gamma$), these processes have excellent sensitivity to off-shell contributions from new particles that could have masses of 100s or 1000s of \tev.

As such, there is a compelling argument for a continued programme of heavy flavour physics in the coming decades.
Rapid progress has been made in the heavy flavour sector since 2000, driven by new experimental facilities including the BaBar and Belle experiments at the PEP-II and KEKB asymmetric $e^+e^-$ $B$ factories, the CDF and D0 experiments at the Tevatron, and the ATLAS, CMS and LHCb experiments at the Large Hadron Collider (LHC).  
This experimental progress has been matched by ongoing developments in theory, including improved calculations of crucial parameters and the introduction of new methods to enable better exploitation of the data.
The Belle~II experiment at the SuperKEKB asymmetric $e^+e^-$ collider and the ATLAS, CMS and LHCb experiments at the LHC and its high luminosity upgrade (HL-LHC) provide exciting prospects for continued reduction in experimental uncertainties, in parallel with continued progress in phenomenological calculations.  

In this document, the expected sensitivities that can be achieved at the different experiments are compared for a selected set of key observables in heavy flavour physics, involving $b$ and $c$ hadrons and $\tau$ leptons.  
The experiments are designed to have different strengths, and therefore some are better suited to certain observables than others --- this provides complementarity in the international heavy flavour physics programme.

Neither the set of observables nor the experiments considered in this document are exhaustive.
In particular, there are several important observables in both kaon and muon physics, and a number of additional observables in charm physics, that are not discussed here and could be investigated at additional facilities.  
Moreover, while this document covers a timescale up to around 20 years from now, in the further future there are proposals for facilities and experiments that could make important contributions for heavy flavour physics.  
Nonetheless, the unique strengths of the current experiments mean that many of the projections in this document are not expected to be surpassed in precision in the foreseeable future.

The remainder of this document is structured as follows.
In Sec.~\ref{sec:experiments} the four experiments are introduced, with brief discussions of their capabilities and operation plans.
The set of observables is introduced in Sec.~\ref{sec:observables}, the uncertainty projections are presented in Sec.~\ref{sec:projections}, and a summary concludes the document in Sec.~\ref{sec:summary}.

\section{The experiments}
\label{sec:experiments}

ATLAS and CMS are general-purpose detectors at the LHC.  
They will operate until the end of Run~3 (2026) at peak instantaneous luminosity of ${\cal L}_{\rm peak} = 2 \times 10^{34}\,\mathrm{cm}^{-2}\,\mathrm{s}^{-1}$.
After upgrades during long shutdown 3 (LS3), they will be able to operate in Run~4 and thereafter at ${\cal L}_{\rm peak} = 7.5 \times 10^{34}\,\mathrm{cm}^{-2}\,\mathrm{s}^{-1}$, corresponding to up to 200 interactions per bunch crossing compared to about 55 during Run~3.
Both experiments expect to record an integrated luminosity of $3000\,\mathrm{fb}^{-1}$, suitable for physics analysis, by the end of HL-LHC operation (2041). 

LHCb is a forward spectrometer at the LHC, optimised for heavy flavour physics.  
It will operate in its current configuration at ${\cal L}_{\rm peak} = 2 \times 10^{33}\,\mathrm{cm}^{-2}\,\mathrm{s}^{-1}$ until the end of Run~4 (2033) by which time it is expected to have recorded $50\,\mathrm{fb}^{-1}$ of high energy $pp$ collision data.  
After an upgrade during LS4, LHCb will operate at ${\cal L}_{\rm peak} = 1.0\text{--}1.5\times10^{34}\,\cm^{-2}\,\sec^{-1}$, corresponding to 28--42 interactions per bunch crossing, and record a sample of at least $300\,\invfb$ by the end of HL-LHC operation.

The Belle~II detector is installed at the SuperKEKB $e^+e^-$ collider, which operates at energies corresponding or near to the mass of the $\Upsilon(4S)$ resonance.
Run 2 data taking resumed in 2024 and will continue until 2032, by when a total integrated luminosity of $10\,\mathrm{ab}^{-1}$ is expected.
An upgrade of the detector is then planned to allow higher instantaneous luminosity to be achieved, resulting in a total sample of $50\,\mathrm{ab}^{-1}$.

\subsection{ATLAS}
\label{sec:atlas}
The ATLAS detector \cite{PERF-2007-01} features a forward–backward symmetric cylindrical geometry and nearly complete solid angle coverage. As a general purpose detector, ATLAS is designed to facilitate the study of a range of particle physics phenomena encompassing both precision measurements of known particles and direct searches for physics beyond the SM. 
This includes significant contributions to $B$-physics research, particularly in investigating rare decays and measuring \CP violation, achieving results that are competitive with those of dedicated experiments.

To maintain optimal performance during the HL-LHC phase, the ATLAS detector will undergo substantial upgrades \cite{ATL-PHYS-PUB-2016-026}. The inner detector will be replaced with the all-silicon Inner Tracker (ITk) \cite{ATLAS:2017svb}, which will extend the pseudorapidity from range from $|\eta| < 2.5$ to $|\eta| \lesssim 4$, and utilize advanced pixel and strip technologies with radiation tolerance up to $10^{16}\,n_{\mathrm{eq}}/\mathrm{cm}^2$. The calorimeter readouts will be enhanced for finer granularity, and the Muon Spectrometer will receive additional resistive plate and thin-gap chambers to manage increased event rates. Additionally, the trigger and data acquisition system will be significantly upgraded, enabling a Level-0 trigger rate of $1\,\mathrm{MHz}$ and achieving a total data throughput of approximately $5\,\mathrm{TB/s}$.
These upgrades are expected to enhance the invariant mass and proper decay time resolution in $B$-physics analyses by approximately $30\%$, facilitating even more precise measurements in flavour physics studies.

\subsection{CMS}
\label{sec:cms}

The CMS detector~\cite{CMS:2024} was not specifically designed for flavour physics, but it demonstrated its competitiveness in flavour physics observables during the first three LHC runs.
Starting in 2018, CMS increased its investment in online resources to expand its flavour physics programme, lowering the threshold on the dimuon trigger and introducing a displaced single-muon trigger~\cite{CMS:parking}. These triggers were designed with two goals: to maximize the acceptance of events with one or more muons in the final state, and to collect a high purity $b\bar{b}$ sample; the selection criteria on the trigger side allow for an unbiased sample of $b$ decays on the probe side of the event, extending CMS sensitivity to fully hadronic final states. 

The HL-LHC upgrade of CMS will bring significant improvements to the muon spectrometer and barrel calorimeter, a complete replacement of the silicon tracker and endcap calorimeter, and the introduction of a new MIP timing detector layer. 
These upgrades will further enhance CMS sensitivity to flavour physics. Muon and track acceptance will be extended with larger coverage, from $|\eta|<2.5$ up to $|\eta|<4$ for tracks;
enhanced timing will provide charged hadron particle identification capabilities in selected $p_{\mathrm{T}}$ ranges. 
The trigger and data acquisition system will be upgraded, enabling a hardware trigger rate of 750 kHz.  
The addition of tracking capabilities in the hardware trigger will allow the design of dedicated triggers on hadronic final states and will enable better purity in muon and electron selection at low $p_{\mathrm{T}}$. 
This will bring both quantitative and qualitative improvements in the trigger decision process.  Based on this, the CMS projections presented in this document are derived under the assumption that the HL-LHC single- and double-muon triggers will be at least as good as those currently operated. It is also assumed that the benefit of the detector upgrades, not entirely accounted for in the extrapolation strategy, will at least compensate for the larger pile-up.

\subsection{LHCb}

The LHCb experiment is designed to exploit the copious production of beauty and charm quarks in the LHC's high energy $pp$ collisions. 
The detector has a forward geometry in the pseudorapidity range $2<\eta<5$, optimised for flavour physics.
It features a vertex locator that surrounds the $pp$ interaction point, to enable precise reconstruction of the displaced vertices that are characteristic of beauty and charm hadron decays.  
Tracking stations up- and downstream of a dipole magnet provide precise momentum resolution, and ring-imaging Cherenkov detectors are used to distinguish charged pions, kaons and protons.
A calorimetry system provides electron, photon and neutral pion identification, while the most downstream active elements of the detector are used to identify muons.

The LHCb detector recorded a data sample corresponding to $9\,\invfb$ of $pp$ collisions during the LHC Runs~1 and~2.
A major upgrade of the detector for Run~3 has been achieved during the second LHC long shutdown~\cite{LHCb-DP-2022-002}.
While the original design featured a hardware trigger that reduced the readout rate to around $1\,\mhz$, resulting in significant inefficiencies for certain decay modes, the upgraded detector is read-out at the full LHC bunch-crossing rate of $40\,\mhz$ with all trigger decisions implemented in software.

By the end of the LHC Run~4, the current LHCb detector will have recorded its design goal of $50\,\invfb$ of integrated luminosity.
A second major upgrade, called LHCb Upgrade~II, is needed to exploit fully the flavour physics potential of the HL-LHC.  
Upgrades to all detector subsystems are foreseen to be able to operate at higher instantaneous luminosity~\cite{LHCb-TDR-023,LHCb-TDR-026}.
A key feature is the addition of timing information in several subdetectors to associate particles to the correct $pp$ interaction vertex and thereby keep backgrounds at levels comparable to or below those in the current detector.

\subsection{Belle~II}
\label{sec:belleii}

Belle~II is a general-purpose detector installed at the SuperKEKB $e^+e^-$ collider, which operates at energies
corresponding or near to the mass of the $\Upsilon(4S)$ resonance. It has a solid-angle coverage close to $4\pi$, which combined with the known initial-state kinematics, provides excellent missing-four-momentum reconstruction. It consists of seven subdetectors that provide excellent vertexing, charged-particle tracking, particle ID and neutral-particle reconstruction. Details of the current Belle~II detector can be found in Ref.~\cite{Belle-II:2010dht}.

Belle~II's Run 1 data-taking period was between 2019 and 2022 prior to a long shutdown,
during which the original pixel detector (that was lacking most of its second layer) was replaced by a complete device. The integrated luminosity collected in Run 1 corresponds to $428\,\mathrm{fb}^{-1}$. 

Run 2 data taking resumed in 2024 and will continue into the 2030s, when the expected integrated luminosity will be approximately $10\,\mathrm{ab}^{-1}$.
At this time, a second long shutdown is scheduled, with the goal of increasing the tolerance of the detector to
machine-related backgrounds, while maintaining the overall performance. The most significant upgrades will be to replace the silicon detectors with a monolithic active pixel silicon detector and make the central drift chamber background tolerant. 
In parallel, the accelerator will carry out an extension of the RF sources and a possible redesign of the interaction region. 
The integrated luminosity recorded by 2042 will be $50\,\mathrm{ab}^{-1}$, if the upgrade of the interaction region takes place, or $30\,\mathrm{ab}^{-1}$ otherwise.

\section{Selected key observables}
\label{sec:observables}

The observables that are considered to illustrate the physics reach of the experiments are described below, separated into groups according to the type of physics they probe.
These observables are in general considered to be theoretically clean, although in some cases theoretical uncertainties associated to interpretation of the measurements will not be negligible relative to the future achievable experimental precision.  
Detailed discussion of these issues goes beyond the scope of this document, but in many cases additional experimental information can be exploited to further constrain theoretical uncertainties.

The Physics Preparatory Group (PPG) of the European Strategy for Particle Physics have prepared a set of benchmark measurements to allow comparison of the physics potential of different projects, in particular future colliders~\cite{ESPPU-benchmarks}.  
This is a different objective than used for the broader list of observables described here, which is intended to give a representative sample of the most important measurements in beauty, charm and tau physics.  
Nonetheless, almost all the benchmark measurements in flavour physics defined by the PPG are included, the exceptions being the $\tau$ lifetime, the branching fractions ${\cal B}(\tau \to \mu \nu\overline{\nu})$ and ${\cal B}(\tau \to e\nu\overline{\nu})$, and CKM elements from on-shell $W$ decays.

\paragraph{CKM angles:}
The usual representation of the CKM unitarity triangle has its three internal angles labelled as $\beta$, $\alpha$ and $\gamma$ (sometimes denoted $\phi_1$, $\phi_2$ and $\phi_3$).
The angle $\beta$ is conventionally determined from the measurement of mixing-induced \CP violation in $B^0 \to \jpsi K_{\rm S}^0$ decays, an approach that has a small theoretical uncertainty due to contributions from subleading amplitudes.  
The angle $\alpha$ is measured exploiting isospin relations between decay rates and \CP violation parameters in $B \to \pi\pi, \rho\pi$ and $\rho\rho$ decays, with very small residual theoretical uncertainty expected due to isospin-breaking effects.
The angle $\gamma$ is evaluated from \CP violation effects in $B^+ \to DK^+$ and similar processes with the neutral $D$ meson reconstructed in a final state that is accessible to both $D^0$ and $\overline{D}{}^0$ decays, where use of experimental inputs on parameters describing the $D$ decays reduces theoretical uncertainties to a negligible level.  

The angles above are related to the triangle formed from the unitarity relation between the rows of the CKM matrix related to couplings of the $b$ and $d$ quarks to up-type quarks.  
For the corresponding triangle formed from the couplings involving the $b$ and $s$ quarks, the equivalent angle to $\beta$, denoted $\beta_s$ or as $\phi_s = -2\beta_s$, is typically determined from $B_s^0 \to \jpsi \phi$ decays.

\paragraph{\boldmath \CP\ violation in loop-dominated $B$ decays:}
The benchmark measurements of CKM angles discussed above involve tree-level $B$ decays.  
Processes that instead are dominated by flavour-changing neutral current loop-level amplitudes (so-called penguin diagrams) are sensitive to possible contributions from physics beyond the SM at high energy scales. 
The parameter of mixing-induced \CP\ violation in $B^0\to \eta^\prime K_{\rm S}$ decays, denoted $S(B^0\to \eta^\prime K_{\rm S}^0)$, is an excellent example of how such models can be tested.
Similarly, the parameters $\phi_s(B^0_s\to \phi\phi)$ and $\phi_s(B^0_s\to K^{*0}\overline{K}{}^{*0})$ can probe for similar contributions in $B_s^0$ decays.

\paragraph{\boldmath \CP\ violation in $B_{(s)}^0\text{--}\overline{B}{}^0_{(s)}$ mixing:}
\CP\ violation in mixing occurs when the mass eigenstates of the $B_{(s)}^0\text{--}\overline{B}{}^0_{(s)}$ system are not equal admixtures of the $B_{(s)}^0$ and $\overline{B}{}^0_{(s)}$ flavour eigenstates.  
Such effects are very small in the SM and therefore can only be determined with decay modes where \CP\ violation in decay is expected to be absent; semileptonic (sl) decays are typically used.  
The parameters of \CP\ violation in mixing in the $B_s^0$ and $B^0$ systems are denoted $a_{\rm sl}^s$ and $a_{\rm sl}^d$, respectively.  

\paragraph{\boldmath \CP\ violation in the charm sector:}
All \CP violation effects in the charm sector are suppressed in the SM, but \CP violation in decay has nonetheless been observed through the observable $\Delta A_{\CP} = A_{\CP}\left(D^0 \to K^+K^-\right)-A_{\CP}\left(D^0 \to \pi^+\pi^-\right)$, where $A_{\CP}$ is the asymmetry between the $D^0 \to f$ and $\overline{D}{}^0 \to f$ decay rates to a final state $f$.
Improved understanding of whether this effect is due to SM dynamics or whether other contributions are involved can be obtained by exploiting flavour symmetries, with $A_\CP(D^+\to \pi^+\pi^0)$ and  $A_\CP(D^0\to \pi^0\pi^0)$ providing examples of important observables.  
Searching for decay-time-dependent \CP violation effects provides further sensitivity to beyond SM dynamics, with the parameters $A_\Gamma$ (determined with $D^0 \to K^+K^-$ and $\pi^+\pi^-$ decays) and $\Delta x$ (with $D^0\to K_{\rm S}^0\pi^+\pi^-$ decays) of particular importance.

\paragraph{\boldmath Semileptonic $B$ decays:}
Improved knowledge of the magnitudes of the CKM matrix elements $\left|V_{ub}\right|$ and $\left|V_{cb}\right|$ is needed to improve the precision of the global CKM fits.  
These can be determined from decays mediated by the semileptonic $b \to u \ell \nu$ and $b \to c \ell \nu$ processes, respectively, where inclusive and exclusive approaches have different theoretical uncertainties.  
The $\left|V_{ub}\right|$ and $\left|V_{cb}\right|$ determinations from exclusive decays are expected to have smaller theory uncertainties, and the use of different $b$-hadron species provides important consistency checks. 
(Decays of on-shell $W$ bosons can also be used to determine magnitudes of CKM matrix elements, but detailed sensitivity studies are not currently available.)

Semileptonic decays can also be used to test the SM prediction of universality between the charged current weak interactions with different lepton flavours.
This can be done through the observables $R(D^{(*)})$, which are the ratios of branching fractions of $B \to D^{(*)}\ell \nu$ decays for final states with $\ell = \tau$ compared to $\ell =\mu~\text{or}~e$.
Current measurements of these quantities indicate some tension with the SM predictions.

\paragraph{\boldmath Leptonic $B$ decays:}
Decays of $B$ mesons to two leptons are suppressed in the SM due to the V$-$A nature of the weak interaction and the small lepton masses.
As the SM contributions are well known, these provide excellent sensitivity to extensions of the SM, in particular models involving new scalar or pseudoscalar mediators.
The branching fractions ${\cal B}(B^0_s\to \mu^+\mu^-)$ and ${\cal B}(B^0 \to \mu^+\mu^-)$ are of great interest as are the effective lifetime $\tau_{\rm eff}(B^0_s\to \mu^+\mu^-)$ and 
the mixing-induced \CP violation parameter $S(B^0_s\to \mu^+\mu^-)$.
For the charged $B$ case it is possible to study both ${\cal B}(B^+\to \tau^+\nu_{\tau})$ and ${\cal B}(B^+\to \mu^+\nu_{\mu})$, providing sensitivity also to new interactions that violate lepton universality.

\paragraph{\boldmath Flavour-changing neutral current $b \to s \ell\ell$ decays:}
Decays mediated by flavour-changing neutral currents are suppressed in the SM and therefore measurements of theoretically clean observables in these processes are of particular interest.
Semileptonic decays offer a broader range of observables compared to leptonic decays, and probe different operators in the effective field theory approach; these therefore have complementary sensitivity to SM extensions.
Recent results for rates and angular observables in $b \to s \ell\ell$ decays are in tension with SM predictions making improved understanding of high importance.  
Among the large number of angular observables that can be measured in $B^0\to K^{*0}\mu^+\mu^-$ decays, the quantity $P_5^\prime$ in the dilepton invariant mass squared ($q^2$) range $[4.0, 6.0]\,\gev^2$ gives a representative example of the  accuracy that can be reached.  
The branching fractions ${\cal B}(B^{+,0}\to K^{+,*0}\nu\overline{\nu})$ and ${\cal B}(B^{+,0}\to K^{+,*0}\tau^+\tau^-)$ are also of significant interest, where for the latter projections are given in terms of the expected upper limit assuming no significant signal is observed.

\paragraph{\boldmath Flavour-changing neutral current $b \to s \gamma$ decays:}
Radiative flavour-changing neutral current processes offer further complementary SM tests.
The inclusive branching fraction ${\cal B}(B\to X_s\gamma)$, where $X_s$ denotes any hadronic final state with unit net strangeness, is highly sensitive to beyond SM contributions to the loop amplitude by which it is mediated. 
A specific feature of $b \to s \gamma$ decays is that the emitted photon is highly polarised in the SM due to the V$-$A nature of the weak interaction.  
The polarisation can be probed in several complementary ways: through mixing-induced \CP violation with the observables $S(B^0 \to K_{\rm S}^0\pi^0\gamma)$ and $S(B^0_s\to \phi\gamma)$, through angular distributions of $B^0\to K^{*0}e^+e^-$ decays at very low $q^2$ values with the observable $A_{\rm T}^{(2)}$, and through the polarisation in decays of baryons with the observable $\alpha_\gamma(\mathit{\Lambda}_b^0\to \mathit{\Lambda}^0\gamma)$.

\paragraph{\boldmath Lepton flavour violation in $\tau$ decays:}
Violation of lepton flavour is known to occur in neutrino oscillations, but charged lepton flavour violation related to this effect is unobservably small in the SM and therefore any measurement of a non-zero effect would be an unambiguous sign of physics beyond the SM.
Equally, upper limits on branching fractions such as ${\cal B}(\tau^+\to \mu^+\gamma)$ and ${\cal B}(\tau^+\to \mu^+\mu^+\mu^-)$ provide stringent constraints on possible extensions of the SM.

\begin{table}[p] 
    \centering
    \renewcommand{\arraystretch}{1.1}
\scalebox{0.82}{
    \begin{tabular}{c| c|c|c|c}
    \hline
    Experiment & \bf{ATLAS} & \bf{CMS} & \bf{LHCb} & \bf{Belle~II} \\
    Assumed data sample & $20.3$-$ 99.7\,\mathrm{fb}^{-1}$ &$116 $-$ 140\,\mathrm{fb}^{-1}$ & $2$-$9\,\mathrm{fb}^{-1}$ & $364$-$1075~\mathrm{fb}^{-1}$ \\
    \hline         
    \multicolumn{5}{l}{\bf CKM angles} \\
    $\beta$ & --- & --- & $0.57\degrees$ \cite{LHCb-PAPER-2023-013} & $1.2^{\circ}$ \cite{Belle-II:2018jsg} \\
    $\alpha$ & --- & --- & --- & $6.6^{\circ}$ \cite{Belle-II:2022cgf} \\
    $\gamma$ & --- & --- & $2.8\degrees$ \cite{LHCb-CONF-2024-004} & $13^{\circ}$ \cite{Belle-II:2022cgf} \\
    $\phi_s$ [mrad] & $42$ \cite{BPHY-2018-01} & $23$ \cite{BPH-23-004} & $20$ \cite{LHCb-PAPER-2023-016} & --- \\
    \hline
    \multicolumn{5}{l}{\bf \boldmath \CP\ violation in loop-dominated decays} \\
    $S(B^0\to \eta^\prime K_{\rm S}^0$) & --- & --- & --- & 0.087 \cite{Belle-II:2022cgf} \\
    $\phi_s(B^0_s\to \phi\phi)$ [mrad] & --- & --- & $69$ \cite{LHCb-PAPER-2023-001} & --- \\
    $\phi_s(B^0_s\to K^{*0}\overline{K}{}^{*0})$ [mrad] & --- & --- & $130$ \cite{LHCb-PAPER-2017-048} & --- \\
    \hline
    \multicolumn{5}{l}{\bf \boldmath \CP\ violation in $B_{(s)}^0\text{--}\overline{B}{}^0_{(s)}$ mixing} \\
    $a_{\rm sl}^s$ [$10^{-4}$] & --- & --- & $33$ \cite{LHCb-PAPER-2016-013} & --- \\
    $a_{\rm sl}^d$ [$10^{-4}$] & --- & --- & $36$ \cite{LHCb-PAPER-2014-053} & $40$ \cite{BaBar:2014bbb} \\
    \hline
    \multicolumn{5}{l}{\bf \boldmath \CP\ violation in the charm sector} \\
    $\Delta A_{\CP}$ [$10^{-5}$] & --- & --- & $29$ \cite{LHCb-PAPER-2019-006} & 630 \cite{Belle-II:2018jsg} \\
    $A_\CP(D^{+,0}\to \pi^{+,0}\pi^0)$ [$10^{-5}$] & --- & --- & 900 \cite{LHCb-PAPER-2021-001},~--- & 870,~750 \\
    $A_\Gamma(KK,\pi\pi)$ [$10^{-5}$] & --- & --- & $11$ \cite{LHCb-PAPER-2020-045} & --- \\
    $\Delta x(D^0\to K_{\rm S}^0\pi^+\pi^-)$ [$10^{-5}$] & --- & --- & $18$ \cite{LHCb-PAPER-2021-009} & $140$ \cite{Belle-II:2024gfc} \\
    \hline
    \multicolumn{5}{l}{\bf \boldmath Semileptonic $B$ decays} \\
    $\left|V_{ub}\right|$ & --- & --- & 6\% \cite{LHCb-PAPER-2015-013} & 6.3\% \cite{Belle-II:2024xwh}\\
    $\left|V_{cb}\right|$ & --- & --- & --- & 1.7\% \cite{Belle:2023xgj}\\
    $R(D)$, $R(D^*)$ & --- & --- & 14\% \cite{LHCb-PAPER-2024-007}, 6\% \cite{LHCb-PAPER-2022-052} & 12\%, 7\% \cite{Belle-II:2022cgf} \\
    \hline
    \multicolumn{5}{l}{\bf \boldmath Leptonic $B$ decays} \\
    ${\cal B}(B^0_s\to \mu^+\mu^-)$ [$10^{-9}$] & $^{+0.8}_{-0.7}$ \cite{BPHY-2018-09} & $0.45$ \cite{BPH-21-006} & $0.48$ \cite{LHCb-PAPER-2021-008} & --- \\
    ${\cal B}(B^0  \to \mu^+\mu^-)$ [$10^{-10}$] & $<2.1^{*}$ \cite{BPHY-2018-09} & $<1.5$ \cite{BPH-21-006} & $0.79$ \cite{LHCb-PAPER-2021-008} & --- \\
    $\tau_{\rm eff}(B^0_s\to \mu^+\mu^-)$ [$\ps$] & $^{+0.45}_{-0.18}$ \cite{BPHY-2020-07} & $0.23$ \cite{BPH-21-006} & $0.29$ \cite{LHCb-PAPER-2021-008} & --- \\
    $S(B^0_s\to \mu^+\mu^-)$ & --- & --- & --- & --- \\
    ${\cal B}(B^+\to \tau^+\nu_{\tau})$ & --- & --- & --- & 34\% \cite{Belle-II:2022cgf} \\
    ${\cal B}(B^+\to \mu^+\nu_{\mu})$ & --- & --- & --- & 41\% \cite{Belle-II:2022cgf} \\
    \hline
    \multicolumn{5}{l}{\bf \boldmath Flavour-changing neutral current $b \to s \ell\ell$ decays} \\
    $P_5^\prime(B^0\to K^{*0}\mu^+\mu^-)$ [$10^{-3}$] ${}^\dagger$ & 390 \cite{BPHY-2013-02} & 100 \cite{BPH-21-002} & 111 \cite{LHCb-PAPER-2020-002} &  --- \\ 
    ${\cal B}(B^{+,0}\to K^{+,*0}\nu\overline{\nu})$ & --- & --- & --- & 57\%, 110\% \cite{Belle-II:2022cgf} \\
    ${\cal B}(B^{+,0}\to K^{+,*0}\tau^+\tau^-)$ [$10^{-4}$] & --- & --- & --- & $<10$, $<18$ \cite{belleIIinprep}\\
    \hline
    \multicolumn{5}{l}{\bf \boldmath Flavour-changing neutral current $b \to s \gamma$ decays} \\
    ${\cal B}(B\to X_s\gamma; E_\gamma > 1.6\,\gev)$ & --- & --- & --- & $(16-18)\%$ \cite{Belle-II:2022cgf} \\
    $S(B^0 \to K_{\rm S}^0\pi^0\gamma)$ & --- & --- & --- & 0.27 \cite{Belle-II:2024uzp} \\
    $S(B^0_s\to \phi\gamma)$ & --- & --- & 0.32 \cite{LHCb-PAPER-2019-015} & --- \\
    $A_{\rm T}^{(2)}(B^0\to K^{*0}e^+e^-; \text{very\ low}\ q^2)$ & --- & --- & 0.10 \cite{LHCb-PAPER-2020-020} & 0.76 \cite{Belle:2024mml} \\
    $\alpha_\gamma(\mathit{\Lambda}_b^0\to \mathit{\Lambda}^0\gamma)$ & --- & --- & 0.26 \cite{LHCb-PAPER-2021-030} & --- \\
    \hline
    \multicolumn{5}{l}{\bf \boldmath Lepton flavour violation in $\tau$ decays} \\
    ${\cal B}(\tau^+\to \mu^+\gamma)$ [$10^{-8}$] & --- & --- & --- & $<7.5$ \cite{Belle-II:2018jsg} \\
    ${\cal B}(\tau^+\to \mu^+\mu^+\mu^-)$ [$10^{-8}$] & $<37.6$ \cite{EXOT-2014-14} & $<2.9$ \cite{BPH-21-005} & $<4.6$ \cite{LHCb-PAPER-2014-052} & $<1.8$ \cite{Belle-II:2024sce} \\
    \hline
    \multicolumn{5}{l}{${}^\dagger$ The sensitivity for the $P_5^\prime$ variable is quoted for the range $q^2 \in [4.0, 6.0]\,\gev^2$ for ATLAS} \\ 
    \multicolumn{5}{l}{~~~and LHCb and $q^2 \in [4.3, 6.0]\,\gev^2$ for CMS.}
    \end{tabular}
}
    \caption{\it 
    Measured uncertainties on selected key observables from the most recent analyses performed by the different experiments. 
    Uncertainties quoted as \% are relative to the measured or SM central value.
    Upper limits, indicated by $<$, are at 90\% confidence level, except where indicated by $^{*}$ in which case the limits are at 95\% confidence level.
    The size of the data samples used varies between the analyses as indicated by luminosity range in the top row. Some Belle~II results include analysis of Belle data.
    } 
    \label{tab:sensitivities-today}
\end{table}

\section{Projections}
\label{sec:projections}

Projected accuracies achievable for relevant subsets of the observables discussed in the previous section have been evaluated for each experiment.
Detailed discussions can be found in Refs.~\cite{ATL-PHYS-PUB-2018-032,ATL-PHYS-PUB-2018-041,ATL-PHYS-PUB-2019-003,ATL-PHYS-PUB-2025-016} for ATLAS, Ref.~\cite{CMS-Proj-2025-xxx} for CMS, Refs.~\cite{LHCb-PII-Physics,LHCb-TDR-023,LHCb-TDR-026} for LHCb and Refs.~\cite{Belle-II:2022cgf,Belle-II:2018jsg} for Belle~II.
For LHCb and Belle~II some projections have been updated based on publications that were completed after the reference documents.

The projections are mostly obtained extrapolating from published results and accounting, where possible, for expected differences in detector performance either due to upgrades or changes in operating conditions. 
Major benefits are expected from improved tracking and vertexing resolution, superior particle identification, higher detector granularity, looser trigger requirements or wider acceptance, depending on the experiment.
Projections have not been performed for every observable in all experiments. In particular, they are not provided for observables for which results have not been published yet. 
This does not imply that such measurements will not be performed or that they will not be precise, but rather that the achievable precision cannot yet be estimated reliably. 
Similarly, potential further improvements in precision due to advances in analysis techniques are not considered, so that in many cases the projections may be considered conservative. 

For the majority of observables, the measurements are expected to remain statistically limited, however where systematic uncertainties are expected to be relevant they are accounted for in the projections. 
The treatment of experimental systematic uncertainties varies between the experiments, but in general the values used in the projections are reduced compared to the current measurements, due to the increased statistics and previously mentioned detector upgrades. 
The expected precision of some observables depends on external inputs, and assumptions about the future uncertainties on these inputs can also differ between the experiments.
Examples include form-factor uncertainties in the determination of $\left|V_{ub}\right|$ and knowledge of fragmentation fractions in ${\cal B}(B^0_s\to \mu^+\mu^-)$.

For some of the projections, ranges are quoted to account for different assumptions in operational conditions such as trigger thresholds or achievable background suppression.  
Where a range is not quoted, nominal values have been assumed but the projections are nevertheless subject to similar uncertainties.

Table~\ref{tab:sensitivities-today} reports the (observed) uncertainties on the observables from the most recent measurements performed by the experiments. 
In many cases, these measurements use only a fraction of the current data collected.   
The sensitivities are projected to the samples expected to be available by the mid-2030s in Table~\ref{tab:sensitivities-2030s} (for LHCb and Belle~II only, since this intermediate point corresponds to notable milestones for those experiments as described in Sec.~\ref{sec:experiments}), and to the expected final data samples of the experiments, to be collected by the early 2040s, in Table~\ref{tab:sensitivities-2040s}.

Specific points to be aware of in the projections are:
\begin{itemize}
    \item $\beta$, $\alpha$ \& $\phi_s$: uncertainties due to subleading SM amplitudes are assumed to remain subdominant, as these can be controlled with measurements of processes related by flavour symmetries.
    \item $\phi_s$: the projected uncertainties are statistical only as systematic uncertainties are expected to be sub-dominant. The ATLAS projection makes use only of opposite-side tagging while the CMS and LHCb projections utilise both same-side and opposite-side tagging.
    \item $\gamma$: auxiliary input measurements on hadronic $D$ decay parameters, as can be obtained for example from the BESIII experiment, are expected to improve in precision and remain subdominant. 
    \item $\left|V_{cb}\right|$ and $\left|V_{ub}\right|$:     Improved lattice QCD form-factor calculations are anticipated, helping to improve precision on both CKM matrix elements. 
    LHCb has published a $\left|V_{cb}\right|$ measurement based on $B_s^0 \to D_s^{(*)-}\mu^+\nu_{\mu}$ decays~\cite{LHCb-PAPER-2019-041}.  This and projections to larger data samples are not included in the tables since it is unclear how the uncertainties will scale in future.
    \item $R(D)$ and $R(D^{\ast})$: Belle~II projections assume that uncertainties related to the $B\to D^{\ast\ast}\ell\nu$ backgrounds are constrained using dedicated measurements. Thus, they are assumed to scale as $1/\sqrt{\mathcal{L}}$.  LHCb assume limiting systematic uncertainties of $3\%$ due to knowledge of backgrounds and normalisation branching fractions.
    \item ${\cal B}(B^0_s\to \mu^+\mu^-)$: uncertainty in the $b$-quark fragmentation probability ratio $f_{s}/f_{d}$ is expected to limit the precision. In the ATLAS projection, a 5.1\% relative uncertainty on $f_{s}/f_{d}$, from the 2021 HFLAV world average~\cite{HFLAV:2019otj}, is used; CMS uses the same uncertainty as in the latest published result~\cite{BPH-21-006} while LHCb assumes this uncertainty can be reduced to $4\%$.
    \item $\mathcal{B}\left(B\to X_{s}\gamma\right)$: the range of uncertainties is determined by whether the uncertainty in the $B$-related background remains 10\% or is improved to 5\% through additional control-sample studies.
    \item $\mathcal{B}\left(\tau^+\to\mu^+\mu^-\mu^+\right)$: For the Belle II projections, the upper limit range indicates scaling of the current expected sensitivity by $1/\sqrt{\mathcal{L}}$ or $1/\mathcal{L}$; further improvements in background suppression are anticipated while maintaining the same efficiency.
    \item The ATLAS projections for $\phi_{s}$, $P_5^\prime(B^0\to K^{*0}\mu^+\mu^-)$ and ${\cal B}(\tau^+\to \mu^+\mu^+\mu^-)$ are based upon analyses of Run 1 (2011-2012) data and therefore should be considered as more conservative than projections based upon more recent analyses. 
    \item The ranges quoted for the ATLAS projections for $\phi_{s}$,  $P_5^\prime(B^0\to K^{*0}\mu^+\mu^-)$ and the $B^0_{(s)}\to \mu^+\mu^-$ observables correspond to different assumptions of HL-LHC trigger thresholds, while for the projection of ${\cal B}(\tau^+\to \mu^+\mu^+\mu^-)$ the range results from different assumptions on the level of background reduction.
    \item The CMS projections are based on the latest Run~2 (2016-2018) analyses; they make assumptions that are different case by case~\cite{CMS-Proj-2025-xxx}.
    CMS has recently demonstrated its capability to study charm \CP violation with a measurement of $A_{\CP}(D^0 \to K^0_{\rm S}K^0_{\rm S})$\cite{BPH-23-005}, however projections are not available for the observables considered here.
\end{itemize}

\begin{table}[p] 
    \centering
    \renewcommand{\arraystretch}{1.1}
\scalebox{0.82}{
    \begin{tabular}{c| c|c}
    \hline
    Experiment & \bf{LHCb} & \bf{Belle~II} \\
    Assumed data sample & $50\,\mathrm{fb}^{-1}$ & $10\,\mathrm{ab}^{-1}$ \\
    \hline         
    \multicolumn{3}{l}{\bf CKM angles} \\
    $\beta$ & $0.20\degrees$ & $0.4^{\circ}$ \\
    $\alpha$ & --- & $2.5^{\circ}$ \\
    $\gamma$ & $0.8\degrees$ & $2.2^{\circ}$ \\
    $\phi_s$ [mrad] & $8$ & --- \\
    \hline
    \multicolumn{3}{l}{\bf \boldmath \CP\ violation in loop-dominated decays} \\
    $S(B^0\to \eta^\prime K_{\rm S}^0$) & --- & 0.023 \\
    $\phi_s(B^0_s\to \phi\phi)$ [mrad] & $22$ & --- \\
    $\phi_s(B^0_s\to K^{*0}\overline{K}{}^{*0})$ [mrad] & $20$ & --- \\
    \hline
    \multicolumn{3}{l}{\bf \boldmath \CP\ violation in $B_{(s)}^0\text{--}\overline{B}{}^0_{(s)}$ mixing} \\
    $a_{\rm sl}^s$ & $7$ & --- \\
    $a_{\rm sl}^d$ & $5$ & $9.5$ \\
    \hline
    \multicolumn{3}{l}{\bf \boldmath \CP\ violation in the charm sector} \\
    $\Delta A_{\CP}$ [$10^{-5}$] & $8$ & 130 \\
    $A_\CP(D^{+,0}\to \pi^{+,0}\pi^0)$ [$10^{-5}$] & 260,~--- & 200,~150 \\
    $A_\Gamma(KK,\pi\pi)$ [$10^{-5}$] & $3.2$ & --- \\
    $\Delta x(D^0\to K_{\rm S}^0\pi^+\pi^-)$ [$10^{-5}$] & $4.1$ & $60$ \\
    \hline
    \multicolumn{3}{l}{\bf \boldmath Semileptonic $B$ decays} \\
    $\left|V_{ub}\right|$ & $2\%$ & $1.8\%$ \\
    $\left|V_{cb}\right|$ & --- & 1.0\% \\
    $R(D)$, $R(D^*)$ & 4.4\%, 3.2\% & 3.0\%, $1.8\%$ \\
    \hline
    \multicolumn{3}{l}{\bf \boldmath Leptonic $B$ decays} \\
    ${\cal B}(B^0_s\to \mu^+\mu^-)$ [$10^{-9}$] & $0.23$ & --- \\
    ${\cal B}(B^0  \to \mu^+\mu^-)$ [$10^{-10}$] & $0.30$ & --- \\
    $\tau_{\rm eff}(B^0_s\to \mu^+\mu^-)$ [$\ps$] & $0.11$ & --- \\
    $S(B^0_s\to \mu^+\mu^-)$ & --- & --- \\
    ${\cal B}(B^+\to \tau^+\nu_{\tau})$ & --- & 10\% \\
    ${\cal B}(B^+\to \mu^+\nu_{\mu})$ & --- & 11\% \\
    \hline
    \multicolumn{3}{l}{\bf \boldmath Flavour-changing neutral current $b \to s \ell\ell$ decays} \\
    $P_5^\prime(B^0\to K^{*0}\mu^+\mu^-)$ [$10^{-3}$] & $29$ & --- \\
    ${\cal B}(B^{+,0}\to K^{+,*0}\nu\overline{\nu})$ & --- & 14\%, 33\% \\
    ${\cal B}(B^{+,0}\to K^{+,*0}\tau^+\tau^-)$ [$10^{-4}$] & --- & $<1.9$, $<3.4$ \\
    \hline
    \multicolumn{3}{l}{\bf \boldmath Flavour-changing neutral current $b \to s \gamma$ decays} \\
    ${\cal B}(B\to X_s\gamma; E_\gamma > 1.6\,\gev)$ & --- & $(6.2-9.6)\%$ \\
    $S(B^0 \to K_{\rm S}^0\pi^0\gamma)$ & --- & 0.07 \\
    $S(B^0_s\to \phi\gamma)$ & 0.062 & --- \\
    $A_{\rm T}^{(2)}(B^0\to K^{*0}e^+e^-; \text{very\ low}\ q^2)$ & 0.043 & 0.15 \\
    $\alpha_\gamma(\mathit{\Lambda}_b^0\to \mathit{\Lambda}^0\gamma)$ & 0.097 & --- \\
    \hline
    \multicolumn{3}{l}{\bf \boldmath Lepton flavour violation in $\tau$ decays} \\
    ${\cal B}(\tau^+\to \mu^+\gamma)$ [$10^{-8}$] & --- & $<1.5$ \\
    ${\cal B}(\tau^+\to \mu^+\mu^+\mu^-)$ [$10^{-8}$] & $<0.64$ & $<(0.08-0.37)$ \\
    \hline
    \multicolumn{3}{l}{${}^\dagger$ The sensitivity for the $P_5^\prime$ variable is quoted for the range $q^2 \in [4.0, 6.0]\,\gev^2$.}
    \end{tabular}
}
    \caption{\it Expected sensitivity to the selected key observables from LHCb and Belle~II with data samples that will be recorded by the early 2030s.
    Uncertainties quoted as \% are relative to the measured or SM central value.
    Upper limits, indicated by $<$, are at 90\% confidence level assuming negligible signal yield.
    }  
    \label{tab:sensitivities-2030s}
\end{table}

\begin{table}[p] 
    \centering
    \renewcommand{\arraystretch}{1.1}
\scalebox{0.82}{
    \begin{tabular}{c| c|c|c|c}
    \hline
    Experiment & \bf{ATLAS} & \bf{CMS} & \bf{LHCb} & \bf{Belle~II} \\
    Assumed data sample & $3000\,\mathrm{fb}^{-1}$ & $3000\,\mathrm{fb}^{-1}$ & $300\,\mathrm{fb}^{-1}$ & $50\,\mathrm{ab}^{-1}$ \\
    \hline         
    \multicolumn{5}{l}{\bf CKM angles} \\
    $\beta$ & --- & --- & $0.08\degrees$ & $0.3^{\circ}$ \\
    $\alpha$ & --- & --- & --- & $0.6^{\circ}$ \\
    $\gamma$ & --- & --- & $0.3\degrees$ & $1.0^{\circ}$ \\
    $\phi_s$ [mrad] & $(4 - 9)$ & $3$ & $3$ & --- \\
    \hline
    \multicolumn{5}{l}{\bf \boldmath \CP\ violation in loop-dominated decays} \\
    $S(B^0\to \eta^\prime K_{\rm S}^0$) & --- & --- & --- & 0.015 \\
    $\phi_s(B^0_s\to \phi\phi)$ [mrad] & --- & --- & $9$ & --- \\
    $\phi_s(B^0_s\to K^{*0}\overline{K}{}^{*0})$ [mrad] & --- & --- & $8$ & --- \\
    \hline
    \multicolumn{5}{l}{\bf \boldmath \CP\ violation in $B_{(s)}^0\text{--}\overline{B}{}^0_{(s)}$ mixing} \\
    $a_{\rm sl}^s$ & --- & --- & $3$ & --- \\
    $a_{\rm sl}^d$ & --- & --- & $2$ & $6.2$ \\
    \hline
    \multicolumn{5}{l}{\bf \boldmath \CP\ violation in the charm sector} \\
    $\Delta A_{\CP}$ [$10^{-5}$] & --- & --- & $3.3$ & 60 \\
    $A_\CP(D^{+,0}\to \pi^{+,0}\pi^0)$ [$10^{-5}$] & --- & --- & 100,~--- & 130,~70 \\
    $A_\Gamma(KK,\pi\pi)$ [$10^{-5}$] & --- & --- & $1.2$ & --- \\
    $\Delta x(D^0\to K_{\rm S}^0\pi^+\pi^-)$ [$10^{-5}$] & --- & --- & $1.6$ & $40$ \\
    \hline
    \multicolumn{5}{l}{\bf \boldmath Semileptonic $B$ decays} \\
    $\left|V_{ub}\right|$ & --- & --- & 1\% & $1.2\%$ \\
    $\left|V_{cb}\right|$ & --- & --- & --- & 1.0\% \\
    $R(D)$, $R(D^*)$ & --- & --- & 3.3\%, 3.0\% & 1.4\%, 1.0\% \\
    \hline
    \multicolumn{5}{l}{\bf \boldmath Leptonic $B$ decays} \\
    ${\cal B}(B^0_s\to \mu^+\mu^-)$ [$10^{-9}$] & $(0.33-0.40)$ & $0.22$ & $0.16$ & --- \\
    ${\cal B}(B^0  \to \mu^+\mu^-)$ [$10^{-10}$] & $(0.32-0.48)$ & $0.12$ & $0.12$ & --- \\
    $\tau_{\rm eff}(B^0_s\to \mu^+\mu^-)$ [$\ps$] & $^{+(0.07-0.11)}_{-(0.05-0.08)}$ & $0.05$ & $0.05$ & --- \\
    $S(B^0_s\to \mu^+\mu^-)$ & --- & --- & 0.2 & --- \\
    ${\cal B}(B^+\to \tau^+\nu_{\tau})$ & --- & --- & --- & 6\% \\
    ${\cal B}(B^+\to \mu^+\nu_{\mu})$ & --- & --- & --- & 5\% \\
    \hline
    \multicolumn{5}{l}{\bf \boldmath Flavour-changing neutral current $b \to s \ell\ell$ decays} \\
    $P_5^\prime(B^0\to K^{*0}\mu^+\mu^-)$ [$10^{-3}$] ${}^\dagger$ & $(47-82)$ & $23$ & $12$ & --- \\
    ${\cal B}(B^{+,0}\to K^{+,*0}\nu\overline{\nu})$ & --- & --- & --- & 8\%, 23\% \\
    ${\cal B}(B^{+,0}\to K^{+,*0}\tau^+\tau^-)$ [$10^{-4}$] & --- & --- & --- & $<0.9$, $<1.5$ \\
    \hline
    \multicolumn{5}{l}{\bf \boldmath Flavour-changing neutral current $b \to s \gamma$ decays} \\
    ${\cal B}(B\to X_s\gamma; E_\gamma > 1.6\,\gev)$ & --- & --- & --- & $(4.7-8.8)\%$ \\
    $S(B^0 \to K_{\rm S}^0\pi^0\gamma)$ & --- & --- & --- & 0.04 \\
    $S(B^0_s\to \phi\gamma)$ & --- & --- & 0.025 & --- \\
    $A_{\rm T}^{(2)}(B^0\to K^{*0}e^+e^-; \text{very\ low}\ q^2)$ & --- & --- & 0.016 & 0.08 \\
    $\alpha_\gamma(\mathit{\Lambda}_b^0\to \mathit{\Lambda}^0\gamma)$ & --- & --- & 0.038 & --- \\
    \hline
    \multicolumn{5}{l}{\bf \boldmath Lepton flavour violation in $\tau$ decays} \\
    ${\cal B}(\tau^+\to \mu^+\gamma)$ [$10^{-8}$] & --- & --- & --- & $<0.7$ \\
    ${\cal B}(\tau^+\to \mu^+\mu^+\mu^-)$ [$10^{-8}$] & $<(0.13-0.64)$ & $<0.39$ & $<0.26$ & $<(0.02-0.17)$ \\
    \hline
     \multicolumn{5}{l}{${}^\dagger$ The sensitivity for the $P_5^\prime$ variable is quoted for the range $q^2 \in [4.0, 6.0]\,\gev^2$ for ATLAS} \\ 
    \multicolumn{5}{l}{~~~and LHCb and $q^2 \in [4.3, 6.0]\,\gev^2$ for CMS.}
    \end{tabular}
}
\caption{\it Sensitivity to the selected key observables from the different experiments with their final data samples, recorded by the early 2040s.
    Uncertainties quoted as \% are relative to the measured or SM central value.
    Upper limits, indicated by $<$, are at 90\% confidence level assuming negligible signal yield.}  
    \label{tab:sensitivities-2040s}
\end{table}

\section{Summary}
\label{sec:summary}

Heavy flavour physics remains a crucial component of the international particle physics programme.
The ATLAS, CMS, LHCb and Belle~II experiments, and their planned upgrades, have complementary strengths but will also compete to achieve the best precision for certain observables, allowing for important consistency checks and even better precision in world average combinations.  

Collectively, these experiments have the capability to significantly advance the precision on essentially all the key observables in $b$, $c$ and $\tau$ physics.
The expected improvement in accuracy that will be achieved by the completion of these experiments in the early 2040s is typically around an order of magnitude from what is available today.
Nevertheless, this is only a partial assessment of the actual physics reach: experimental progress and further analysis optimization could in many cases push the physics reach beyond the values projected here, and more measurements than have been  presented will be performed.
In particular, measurements of many suppressed processes which have not yet been studied and more detailed investigations of differential distributions will be possible with much larger samples that will become available.

The precision that can be achieved with these experiments provides exciting and unprecedented capability to probe the flavour sector of the Standard Model.
The results can be used to constrain or rule out specific models beyond the SM and to assess potential contributions from a wide range of operators, associated with flavour-changing interactions, in effective field theory approaches.
The vastly improved knowledge that will be obtained in this sector could allow the observation of imprints of physics at higher energy scales than can be accessed directly.

\addcontentsline{toc}{section}{References}
\bibliographystyle{LHCb} 
\bibliography{references}

\end{document}